\documentclass[twocolumn,preprintnumbers,amsmath,amssymb]{revtex4}

\usepackage{graphicx}
\usepackage{dcolumn}
\usepackage{bm}
\usepackage{txfonts}
\usepackage{lscape}
\begin{document}

\title{Functional renormalization group for d-wave superconductivity in Hubbard type models}

\author{H. C. Krahl}  
\author{C. Wetterich}
\affiliation{Institut f\"ur Theoretische Physik, 
Universit\"at Heidelberg, Philosophenweg
16, D-69120 Heidelberg, Germany}


\begin{abstract}
The temperature dependence of d-wave superconducting order for two
dimensional fermions with d-wave attraction is investigated by means of the
functional renormalization group with partial bosonization. Below the critical
temperature $T_c$ we find superconductivity, a gap in the electron
propagator and a temperature dependent anomalous dimension. At $T_c$
the renormalized ``superfluid density'' jumps and the approach to $T_c$ from
above is characterized by essential scaling. These features are
characteristic for a phase transition of the Kosterlitz-Thouless (KT) type.

\end{abstract}

\maketitle
The two-dimensional Hubbard model \cite{hubbard} is often associated with
high temperature superconductivity \cite{anderson}. Its ground state is
assumed to be a d-wave superconductor for an appropriate range of doping
\cite{scalapino,tsuei}.
Under the hypothesis of d-wave superconductivity we study within a simplified
effective model the phase transition from the
high temperature ``symmetric phase'' to the low temperature
superconducting phase by means of the functional renormalization
group. As characteristic features we find (i) essential scaling as the
critical temperature $T_c$ is approached from above, (ii) a massless
Goldstone mode for $T<T_c$ leading to superfluidity, (iii) a
temperature dependent anomalous dimension $\eta$ for $T<T_c$, (iv) a
jump in the renormalized superfluid density at $T_c$ and (v) a gap in
the electron propagator for $T<T_c$ which vanishes non-analytically as
$T\rightarrow T_c$.

The features (i)-(iv) are characteristic for a KT
phase transition \cite{kosterlitzthouless73} which is a natural
candidate for a universality class with a global $U(1)$
symmetry. Actually, while in the infinite volume limit the ``bare''
order parameter vanishes in accordance with the Mermin-Wagner theorem,
a renormalized order parameter \cite{wetterichZ} spontaneously breaks
the $U(1)$ symmetry for $T<T_c$, leading to an infinite correlation
length for the Goldstone mode. Vortices arise as topological defects
for $T<T_c$. In the Hubbard model the fermionic excitations (single
electrons or holes) are an important ingredient for driving the
transition.  Finally, coupling our model to electromagnetic fields the
photon acquires a mass through the Higgs mechanism while the Goldstone
mode disappears from the spectrum, being a gauge degree of
freedom. Superfluidity is replaced by superconductivity. Recently, various 
experimental data have been interpreted as signatures of 
KT phase fluctuations \cite{rufenacht,corson,wang,lie}. The direct relation
between real high temperature superconductors and the two-dimensional
Hubbard model is difficult, however.

The problem of d-wave superconductivity in the Hubbard model can be
separated into two qualitative steps in the renormalization flow. In
the first step the fluctuations with momenta
$|\mathbf{p}^2-\mathbf{p}^2_F|>\Lambda^2$ (with $\mathbf{p}_F$ on the
Fermi-surface) have to generate a strong effective interaction in the
d-wave channel.
This phenomenon has indeed been found in renormalization group
studies \cite{zanchi1,zanchi2,halbothmetzner,halbothmetzner2,salmhofer,honerkamp01,honerkampsalmhofer01,honerkamp02,katanin}, the d-wave coupling being triggered by spin
wave fluctuations. It also has been found in various other approaches, see e.g.
\cite{miyake,loh,bickers}.
The second step, involving momenta $|\mathbf{p}^2-\mathbf{p}^2_F|<\Lambda^2$
for the fermions and $\mathbf{p}^2<\Lambda^2$ for bosonic bound states,
has to deal with a situation where the d-wave channel coupling dominates.
We only address here the second step.
We find that the universal behavior in the vicinity of the
critical temperature is dominated by effective bosonic fluctuations
and becomes independent of the microscopic details. This justifies our
approach with a single coupling $\lambda_d$ at the scale $\Lambda$ in
the d-wave channel, even though this remains, of course, only an
approximation. The prize to pay is that $\lambda_d$ is only an
effective coupling -- its relation to the parameters $t$ and $U$ of the
Hubbard model is expected to depend on $T$ and the chemical potential
$\mu$.

We emphasize that we obtain a KT-like phase transition within a purely
fermionic model. The effective bosonic degrees of freedom arise as composite
fields -- we do not start from a bosonic effective theory. The transition
from a fermionic to a bosonic description is a result of the renormalization
flow. The complete decoupling of the fermionic degrees of freedom for the
description of the critical behavior is a non-trivial result which only holds
for a sufficiently large critical temperature $T_c$. (For $T_c\rightarrow 0$
interesting modifications reflect the different universality class of a
quantum phase transition.) Several new features are directly related to the
dominance of fermionic fluctuations sufficiently away from the critical
region: We find that the critical temperature $T_c$ is below a 
``pseudo-critical temperature'' $T_{pc}$ where the effective interaction
strength for the electrons diverges. On the other hand, the temperature range
$T_c<T<T_{pc}$ is strongly affected by fluctuations of composite bosons.
For the transition to d-wave superconductivity we find that $T_c$ is actually
rather near to $T_{pc}$. This contrasts to the transition to
antiferromagnetism for $\mu=0$ where $T_c$ has been found substantially below
$T_{pc}$ \cite{bbw04,bbw05}. Other fermionic aspects concern the relation between
the gap in the electron propagator and the superfluid order. We also find
a quantitatively important finite size effect which modifies the essential
scaling for $T\rightarrow T_c$. Again it is dominated by the fermionic
fluctuations.

Our starting point is an action with nearest and next-nearest
neighbor hopping on a two dimensional square lattice with an
attractive interaction in the d-wave channel
\begin{eqnarray}\label{eq:fermaction}
\Gamma^F_{\Lambda}[\psi]&=&\sum_Q\psi^{\dagger}(Q)(i\omega_F
+\epsilon_{\mathbf{q}}-\mu)\psi(Q)
\\
&-&\lambda_d\sum_{K_1,\dots,K_4}\delta_{K_1+K_2\,,K_3+K_4}
d((\mathbf{k_1-k_2})/2)
\nonumber\\
& &\times d((\mathbf{k_3-k_4})/2)\psi^*_{\downarrow}(K_3)
\psi^*_{\uparrow}(K_4)\psi_{\uparrow}(K_2)\psi_{\downarrow}(K_1)\nonumber
\,.\end{eqnarray}
We employ the short-hand notation $Q=(\omega_F,\mathbf{q})$,
$\omega_F=(2n+1)\pi T$, $n\in \mathbb{Z}$,
$\sum_Q=T\sum_{n}\int_{-\pi}^{\pi}\frac{d^2q}{(2\pi)^2}$. The Fermi
surface is specified by $\epsilon_{\mathbf{q}}=-2t(\cos q_1+\cos
q_2)-4t'\cos q_1\cos q_2=\mu$ and $d(\mathbf{q})=\cos q_1-\cos q_2$ is
a d-wave form factor.  All lengths are measured in units of the
lattice distance which we set to unity. We use partial bosonization
with a bosonic field $\phi$ standing for a fermion bilinear
$\sim\psi\psi$ in the d-wave channel. For the integration of the
fluctuations we employ the concept of the effective average action
$\Gamma_k$ \cite{berges_review02}, which amounts to the quantum
effective action (generating 1PI-vertices) in presence of an infrared
cutoff $k$. It interpolates between the microscopic action
$\Gamma_{k\rightarrow\infty}\approx S$ and the full quantum effective
action $\Gamma_{k\rightarrow 0}=\Gamma$.

The flow of the effective average action obeys an exact
renormalization group equation \cite{cw93},
\begin{eqnarray}\label{eq:floweq}
\partial_k\Gamma_k[\phi,\psi]=\frac{1}{2}\mathrm{STr}
\left\{\tilde{\partial}_k\ln\big[\Gamma^{(2)}_k[\phi,\psi]+R_k\big]\right\}
\,,\end{eqnarray}
where $R_k$ denotes an infrared cutoff and
$\tilde{\partial}_k=\sum_i(\partial_k R^i_k) \frac{\partial}{\partial
R^i_k}$ (the index i refers to cutoffs for fermions and bosons).  The
``supertrace'' $\mathrm{STr}$ runs over field type, momentum and
internal indices, and has an additional minus sign for fermionic
entries. The functional differential equation (\ref{eq:floweq})
involves the logarithm of the full inverse propagator
$\Gamma^{(2)}_k[\phi,\psi]$ in presence of background fields (second
functional derivative of $\Gamma_k$), regulated by
$R_k$. Approximations to the solution of eq. (\ref{eq:floweq}) proceed
by a truncation of $\Gamma_k$ on the r.h.s. with a suitable ansatz.

Our truncation for the effective average action is
\begin{eqnarray}\label{eq:ansatz}
\Gamma_k&=&\sum_Q\psi^{\dagger}(Q)(i\omega_F+\epsilon_{\mathbf{q}}-
\mu)\psi(Q)\\
&+&\sum_Q\phi^*(Q)(iZ\omega_B+A\,[\mathbf{q}]^2t^2)\phi(Q)+\sum_X
U(\phi)\nonumber\\ &-&\sum_{K,Q,Q'}\delta_{K\,,Q+Q'}d((\mathbf{q-q'}
)/2)\nonumber\\ &
&\times\big(\phi^*(K)\psi_{\uparrow}(Q)\psi_{\downarrow}(Q')
-\phi(K)\psi^*_{\uparrow}(Q)\psi^*_{\downarrow}(Q')\big)\nonumber
\,,\end{eqnarray}
with $X=(\tau,\mathbf{x})$,
$\sum_X=\int_{0}^{\frac{1}{T}}d\tau\sum_{\mathbf{x}}$ and for bosons
$Q=(\omega_B,\mathbf{q})$, $\omega_B=2m\pi T$, $m\in
\mathbb{Z}$. Further $[\mathbf{q}]^2$ is defined as
$[\mathbf{q}]^2=q_1^2+q_2^2$ for $q_i\in [-\pi,\pi]$ and continued
periodically otherwise.  We approximate the effective potential by a
quartic polynomial
\begin{eqnarray}\label{eq:effpot}
U(\phi)=\left\{\begin{array}{cc}
\bar{m}^2\delta+\frac{\bar\lambda_{\phi}}{2}\delta^2 & (\mathrm{SYM})\\
\frac{\bar\lambda_{\phi}}{2}(\delta-\delta_0)^2 & (\mathrm{SSB})
\end{array}\right.
\,,\end{eqnarray}
with $\delta=\phi^*\phi$. Here (SYM) is used as long as the minimum of
$U$ is located at $\phi=0$, while the regime with spontaneous symmetry
breaking (SSB) is characterized by non vanishing
$\delta_0=\phi^*_0\phi_0$. The order parameter $\phi_0$ determines the
size of the gap in the fermion propagator
$\Delta(\mathbf{q})=|d(\mathbf{q})\phi_0|$. The generalized couplings
$Z$, $A$, $\bar{m}^2$, $\delta_0$ and $\bar{\lambda}_{\phi}$ depend on
$k$. At the scale $\Lambda$ our truncation is equivalent to the fermionic 
action (\ref{eq:fermaction}) provided
\begin{eqnarray} \label{eq:initialvalues}
(\bar{m}/t)^2=\frac{1}{\lambda_d/t}\,,\quad\bar\lambda_{\phi}=0\,,\quad Z=A=0
\,,\end{eqnarray}
as can be seen by solving for $\phi$ as a functional of $\psi$.

We will follow the flow until a physical scale $k_{ph}$ where
$k^{-1}_{ph}$ corresponds to the macroscopic size of the experimental
probe. (In practice, our choice $k_{ph}/t=10^{-9}=e^{-20.723}$
corresponds to a probe size of roughly 1 cm.) For $T<T_c$ we find a
non vanishing superfluid density $\delta_0(k_{ph})$ and therefore a non
vanishing gap $\Delta$, while $\delta_0(k_{ph}\rightarrow 0)=0$. 

In addition to the truncation we have to specify the regulator functions $R_k$.
For the fermion fluctuations we use a
temperature like regulator,
\begin{eqnarray}
R_k^F(Q)=i\omega_F\cdot\left(\frac{T_k}{T}-1\right)\,,\quad\mathrm{with}\quad
T_k^4=T^4+k^4
\,.\end{eqnarray}
For $k\gtrsim T$ the cutoff $R_k^F$ in the inverse fermion propagator
suppresses the contribution of all fluctuations with momenta
$|\mathbf{p}-\mathbf{p}_F|^2<(\pi k)^2$, even for $T=0$. It becomes
ineffective for $k\ll T$ where no cutoff is needed anymore.  The
bosonic cutoff regularizes the fluctuations of long
range bosonic modes. We use a ``linear cutoff'' for the space like
momenta \cite{litim1,litim2}
\begin{eqnarray}
R_k^B(Q)=A\cdot\left(k^2-[\mathbf{q}]^2t^2\right)\Theta(k^2-[\mathbf{q}]^2t^2)
\,.\end{eqnarray}

The flow equations for the couplings are obtained by projection
from (\ref{eq:floweq}). In (SYM) the effective potential flows as
\begin{eqnarray}\label{eq:U}
\partial_{k}U_k(\delta)&=&\frac{2T(\partial_{k}T_k)}{T_k}\int_{-\pi}^{\pi}
\frac{d^2q}{(2 \pi)^2}
\gamma_\delta\tanh\gamma_\delta
\\
& &\hspace{-1.5cm}+
\sum_{Q}\frac{\partial_kR_k^{\phi}(Q)[P_k^B(-Q)+\bar{m}^2+2
\bar{\lambda}_{\phi}\delta]}
{[P_k^B(Q)+\bar{m}^2+2\bar{\lambda}_{\phi}\delta][P_k^B(-Q)+
\bar{m}^2+2\bar{\lambda}_{\phi}\delta]-\bar{\lambda}_{\phi}^2\delta^2}\nonumber
\,,\end{eqnarray}
where
\begin{eqnarray}
\gamma_\delta&=&\frac{1}{2T_k}\sqrt{(\epsilon_{\mathbf{q}}-\mu)^2
+\Delta^2(\mathbf{q})}\,,\nonumber\\
P_k^B(Q)&=&iZ\omega_B+A\,[\mathbf{q}]^2t^2+R_k^B(Q)
\,.\end{eqnarray}
For the mass parameter and the bosonic quartic coupling this yields
\begin{eqnarray}
\partial_k\bar{m}^2=\frac{\partial}{\partial\delta}\partial_{k}U_k(\delta)|_{\delta=0}\,,\quad\partial_k\bar\lambda_{\phi}=\frac{\partial^2}{\partial^2\delta}\partial_{k}U_k(\delta)|_{\delta=0}
\,.\end{eqnarray}

The fermion fluctuations generate kinetic and gradient terms for the
bound state bosons (SYM regime)
\begin{eqnarray}
\partial_k Z\!\!&=&\!\!-T(\partial_kT_k)\int_{-\pi}^{\pi}\frac{d^2q}{(2\pi)^2}
d(\mathbf{q})^2\frac{\partial}{\partial T_k}\frac{\partial}{i\partial\omega}\\
& &\hspace{2.5cm}\times
\frac{\tanh\frac{\epsilon_{\mathbf{q}}-\mu}{2T_k}+\tanh(\frac{\epsilon_{\mathbf{q}}-\mu}{2T_k}+\frac{i\omega}{2T})}{T^2_k(\frac{\epsilon_{\mathbf{q}}-\mu}{T_k}+\frac{i\omega}{2T})}\Bigg{|}_{\omega=0}\nonumber
\,,\end{eqnarray}
and
\begin{eqnarray}
\partial_kA\!\!&=&\!\!-2T(\partial_kT_k)\frac{1}{t^2}
\frac{\partial}{\partial T_k}\frac{\partial}{\partial (l^2)}
\int_{-\pi}^{\pi}\!\frac{d^2q}{(2\pi)^2}d(\mathbf{q}-\frac{l}{2}\mathbf{e_1})^2\\
& &\hspace{2.5cm}\times
\frac{\tanh(\frac{\epsilon_{\mathbf{q}}-\mu}{2T_k})+\tanh(\frac{\epsilon_{l\mathbf{e_1-q}}-\mu}{2T_k})}{T_k(\epsilon_{\mathbf{q}}+\epsilon_{l\mathbf{e_1-q}}-2\mu)}\Bigg{|}_{l=0}\nonumber
\,.\end{eqnarray}
There are no bosonic contributions to the flow of $Z$ and $A$ to this
level of approximation. The flow starts in the symmetric regime
(\ref{eq:effpot}) with initial values (\ref{eq:initialvalues}) and we
choose $\Lambda/t=0.5$.  For temperatures below a pseudo-critical
temperature $T_{pc}$ the scale dependent mass term $\bar{m}^2$
vanishes at a certain critical scale $k_{SSB}$. This signals local
ordering and indeed corresponds to a divergent effective four fermion
coupling $\bar{\lambda}_d(k)$. Below $k_{SSB}$ we use the truncation
(SSB) in (\ref{eq:effpot}).

\begin{figure}[t]
\begin{center}
\begin{picture}(185,155)(20,0)
\put(194,15){\line(0,1){131}}
\put(215,100.5){$\kappa_*\rightarrow$}
\put(190,150){$k_{ph}$}
\put(-5,76){$\kappa$}
\put(110,-5){$-\ln k/t$}
\put(150,75){$T=T_c$}
\put(150,127){$T<T_c$}
\put(55,35){$T>T_c$}
\includegraphics[width=85mm,angle=0.]{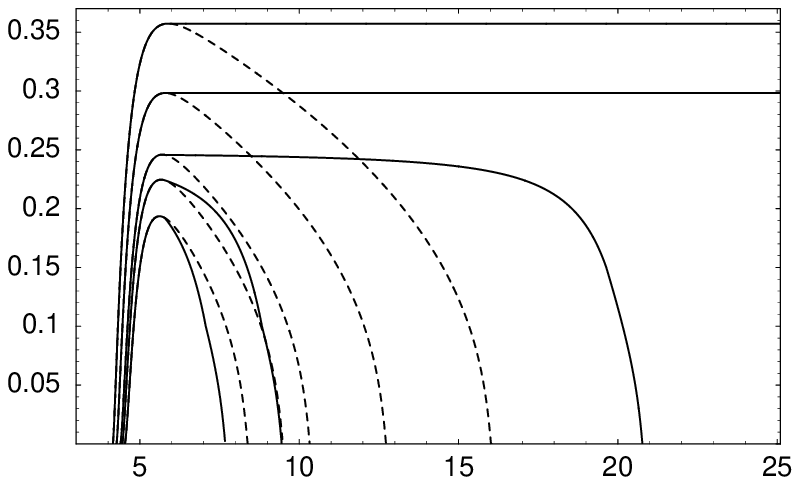}
\end{picture}
\end{center}
\caption{Flow of $\kappa(k)$ for $T/t$=(0.037,\,0.0369\,,$T_c$/t,\,0.0365,\,0.036) with $T_c/t=0.03681$ and $t'/t=-0.1$,\,$\lambda_d/t=5/8$, $\mu/t=-0.6$.
}
\label{figmu02_kappad}
\end{figure}

We do not consider here the region of very small $T$ where
interesting quantum critical phenomena may occur \cite{vojta}. Then classical
statistics dominates the flow for $k\ll T$. A typical flow in the SSB
regime is depicted in fig. \ref{figmu02_kappad}, where we show
$\kappa=t^2 A\delta_0/T$ for various values of $T$ and $t'/t=-0.1$,
$\lambda_d/t=5/8$, $\mu/t=-0.6$. The fermion fluctuations (first term
in eq. (\ref{eq:U})) first drive $\bar{m}^2(k)$ to zero and induce non
vanishing $\kappa(k)$, starting from $\kappa(k_{SSB})=0$. For $k\ll
k_{SSB}$ the effect of the fermion fluctuations is strongly reduced
due to the suppression with powers $k/T$ from $\partial_k T_k$ and the
presence of a non vanishing gap $\Delta$. Now the bosonic fluctuations
take over and tend to reduce the value of $\kappa(k)$ as $k$ is
lowered further. For the region around and on the right of the maximum
of $\kappa(k)$ in fig. \ref{figmu02_kappad} we have already $k\ll
T$. Then the Matsubara frequency $m=0$ dominates the bosonic
contributions and the system gets reduced to classical statistics for
the $O(2)$ linear $\sigma$-model. One can see this effective reduction
from $2+1$ to 2 dimensions explicitly by evaluating the bosonic
contribution in eq. (\ref{eq:U}) (the second term, denoted by
$\partial_kU_k^B$). Keeping only the $m=0$ contribution the flow
equation simplifies considerably
\begin{eqnarray}
\partial_{k}U_k^{B}(\delta)\!\!&=&\!\!\frac{TAk}{2}\int_{-\pi}^{\pi}\frac{d^2q}{(2 \pi)^2}\big(2-\eta(k)(1-\frac{\mathbf{q}^2t^2}{k^2}))\Theta(k^2-\mathbf{q}^2t^2\big)
\nonumber\\
& &\times\Bigg\{\frac{1}{Ak^2+(\delta-\delta_0)\bar\lambda_{\phi}}+\frac{1}{Ak^2+(3\delta-\delta_0)\bar\lambda_{\phi}}\Bigg\}
\,.\end{eqnarray}
Here $\eta(k)=-\frac{\partial\ln A}{\partial\ln k}$ defines the
anomalous dimension.

For a better understanding of the scaling behavior we introduce
rescaled and renormalized quantities
\begin{eqnarray}
u=\frac{t^2}{Tk^2}U\,,\quad\tilde\delta=\frac{t^2A}{T}\delta\,,
\quad\kappa=\frac{t^2A}{T}\delta_0
\,.\end{eqnarray}
This yields for the flow of the rescaled effective potential, now at
fixed $\tilde\delta$ instead of fixed $\delta$,
\begin{eqnarray}
k\partial_ku_k=-2u_k+\eta(k)\,\tilde\delta\,\frac{\partial u_k}{\partial\tilde\delta}+\frac{t^2}{Tk^2}k\partial_kU_k|_{_{\delta}}
\,,\end{eqnarray}
implying for the renormalized field expectation value $\kappa$ and bosonic quartic coupling $\lambda_{\phi}$
\begin{eqnarray}
\partial_k\kappa=-\frac{1}{\lambda_{\phi}}\frac{\partial}{\partial\tilde\delta}\partial_{k}u_k|_{_{\tilde\delta=\kappa}}\,,\quad\partial_k\lambda_{\phi}=\frac{\partial^2}{\partial^2\tilde\delta}\partial_{k}u_k|_{_{\tilde\delta=\kappa}}
\,.\end{eqnarray}
The bosonic contribution to the anomalous dimension in SSB is the same as for the linear $O(2)$ model in two dimensions
\begin{eqnarray}
\eta(k)=\frac{1}{\pi}\frac{\lambda_{\phi}^2\kappa}{(1+2\lambda_{\phi}\kappa)^2}=\frac{1}{4\pi\kappa}+O(\kappa^{-2})
\,,\end{eqnarray}
while the fermionic contribution can be neglected. 

\begin{figure}[t]
\begin{center}
\begin{picture}(185,155)(20,0)
\put(-10,75){$\ln\frac{m_R}{k_{ph}}$}
\put(80,-10){$\sqrt{T_c/(C(k_{ph})+T-T_c)}$}
\includegraphics[width=80mm,angle=0.]{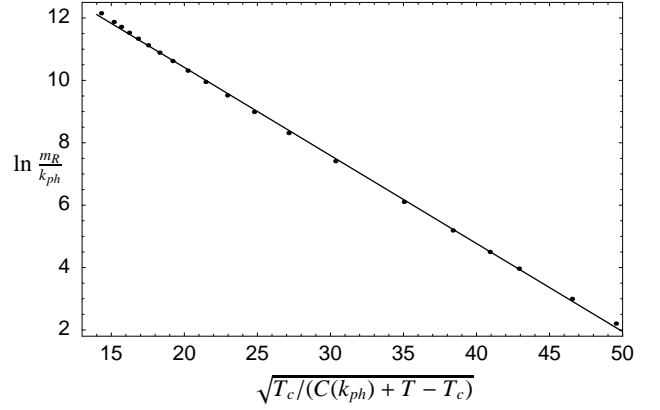}
\end{picture}
\end{center}
\caption{Essential scaling for $T>T_c$ with $C(k_{ph})/t=10^{-5}$. The dots are the numerical solution to the flow equation.
}
\label{figessscalmRT}
\end{figure}

The $\beta$-functions can be expanded in powers of $(2\lambda_{\phi}\kappa)^{-1}$. In particular, to leading order one finds
\begin{eqnarray}\label{eq:betakl}
k\partial_k\kappa=\beta_{\kappa}=O(\kappa^{-1})\,,\quad k\partial_k\lambda_{\phi}=-2\lambda_{\phi}+\frac{\lambda^2_{\phi}}{2\pi}+O(\kappa^{-1})
\,,\end{eqnarray}
resulting in an infrared attractive fixed point for
$\lambda_{\phi}$. The running of $\kappa$ can be mapped to the running
of the coupling in the non-linear $\sigma$-model for arbitrary
$O(N)$-models \cite{wetterichZ,graeter95,vgersdorff}. For $N=2$ also
the term $\sim\kappa^{-1}$ in $\beta_{\kappa}$ has to vanish, even
though this is not seen in the truncation (\ref{eq:ansatz}). The flow
of the classical linear $O(2)$-model has already been studied with
much more extended truncations \cite{vgersdorff}, taking into account
an arbitrary $\tilde\delta$-dependence of $u(\tilde\delta)$ and
arbitrary $\tilde\delta$-dependent wave function renormalizations,
e.g. $A(\tilde\delta)$. All these functions flow quickly to a scaling
form which depends only on one parameter $\kappa$. Evaluated for the
scaling solution one finds to a very good approximation for
$\kappa>\kappa_{KT}$
\begin{eqnarray}\label{eq:impbeta}
k\partial_k\kappa=\beta_{\kappa}=\alpha |\kappa_*-\kappa|^{\frac{3}{2}}\Theta(\kappa_*-\kappa)
\end{eqnarray}
with $\kappa_*=0.248$ and $\alpha=2.54$ \cite{vgersdorff}. We will use
this improved truncation for $\beta_{\kappa}$ for $k<T/12$ for the
region $\kappa>\kappa_{KT}=0.1$. This results in the solid lines in
fig. \ref{figmu02_kappad}, whereas the truncation (\ref{eq:ansatz})
gives the dashed lines.

We define the critical temperature $T_c$ as the temperature for which
the field expectation value $\kappa$ just vanishes at the physical
scale $k_{ph}/t=10^{-9}$, see fig. \ref{figmu02_kappad}. For $T<T_c$
one finds a superfluid condensate in a probe of macroscopic size of
roughly 1 cm. The dependence of $T_c$ on the specific choice of
$k_{ph}$ is extremely weak. For our parameters $t=0.3$ eV corresponds
to $T_c=128$ K while for alternative $\lambda_d/t=1/2,\,1$ we find 
$T_c=58.3,\,366$ K, respectively.
We also note that $T_c$ is only mildly affected by the
improvement of the truncation (\ref{eq:impbeta}).

A specific feature of the KT phase transition is
essential scaling for the temperature dependence of the correlation
length $\xi=m_R^{-1}=(\bar{m}^2/A)^{-\frac{1}{2}}$ just above $T_c$,
\begin{eqnarray}\label{eq:essscal}
m_R=ce^{-\frac{b}{\sqrt{C(k_{ph})+T-T_c}}}
\,.\end{eqnarray}
The finite size correction
$C(k_{ph})\sim(1+\frac{\alpha}{2}\kappa_*^{\frac{1}{2}}
\ln\frac{\Lambda}{k_{ph}})^{-2}$ vanishes logarithmically 
for $k_{ph}\rightarrow 0$. This follows from eq. (\ref{eq:impbeta}).
Close to the critical temperature the omission of the finite size correction
(i.e. $C(k_{ph}=0$) yields to a substantial derivation from the straight line
in fig. 2.

For $T>T_c$ one finds that
$\kappa(k)$ reaches zero at a scale $k_{SR}>k_{ph}$. Continuing the
flow in the symmetric regime for $k_{SR}>k>k_{ph}$ yields $m_R=0.50\,k_{SR}$. We show $m_R(T)$ in fig. \ref{figessscalmRT} (same $t'$,
$\lambda_d$, $\mu$ in all figs.) and find that essential scaling is
smoothened due to $k_{ph}>0$.

\begin{figure}[t]
\begin{center}
\begin{picture}(185,155)(20,-5)
\put(120,105){$100\times\Delta/t$}
\put(83,75){$\kappa$}
\put(110,-5){$T/T_c$}
\includegraphics[width=80mm,angle=0.]{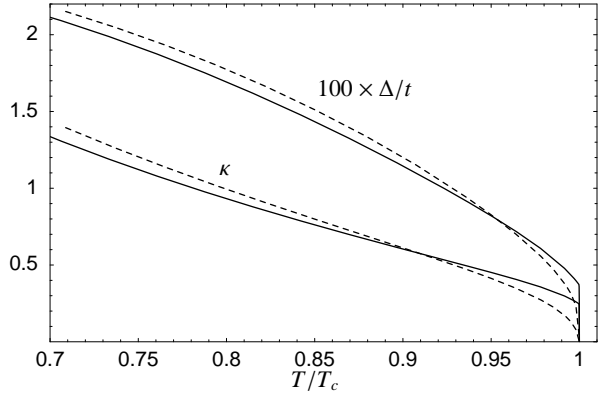}
\end{picture}
\vspace{-5mm}
\end{center}
\caption{Temperature dependence of d-wave superconducting order. We plot the renormalized superfluid density $\kappa$ and the fermionic gap $\Delta(0,\frac{\pi}{2})/t$. Solid/dashed lines are based on eq. (\ref{eq:impbeta})/(\ref{eq:ansatz}).
}
\label{figkappagap}
\end{figure}

From fig. 1 we can also extrapolate the pseudo-critical temperature $T_{pc}$.
This corresponds to the temperature where $\kappa(k)$ bends over immediately
after becoming non-zero at $k=k_{SSB}$ and is driven to zero again more or
less at the same scale $k_{SR}\approx k_{SSB}$. In other words, $T_{pc}$
corresponds to the highest temperature for which the initial flow in the
symmetric regime hits a vanishing mass term $\bar{m}^2$ and therefore a
divergent effective four fermion vertex $\sim \bar{m}^{-2}$. Often this
temperature is associated with the critical temperature. We find
$T_{pc}/t=0.0372$, only slightly above the value  of the critical temperature
$T_{c}/t=0.0368$.

It is characteristic for two dimensions that the renormalized order
parameter $\kappa$ jumps at $T_c$ even though the transition shows
scaling behavior. (Strictly speaking, this discontinuity only occurs
for $k_{ph}\rightarrow 0$ whereas it is smoothened for nonzero
$k_{ph}$.) The scaling solution at $T_c$ corresponds to a fixed point
$\kappa_*$ where $\beta_{\kappa}(\kappa_*)=0$ with
$\beta_{\kappa}(\kappa<\kappa_*)>0$. In the region
$\kappa(k)<\kappa_*$ the flow drives $\kappa$ to zero and the system
is in the symmetric phase. In contrast, for $\kappa>\kappa_*$ the flow
can never cross the fixed point. Therefor $\kappa(k_{ph})$ remains
larger than $\kappa_*$, resulting in a minimal value for $\kappa$ in
the ordered phase. For $k_{ph}\rightarrow 0$ this results in a jump
$\Delta\kappa=\kappa_*$ between the two phases. In superfluids
$\kappa(k_{ph})$ determines the renormalized superfluid density. The
value resulting in the vortex picture \cite{kosterlitzthouless73} is
$\kappa_*=\frac{1}{\pi}$. Our truncation (\ref{eq:impbeta}) is already
parameterized in terms of $\kappa_*$. In fig. \ref{figkappagap} we
plot the value of the superfluid density $\kappa(k_{ph})$ as well as
the gap in the electron propagator $\Delta(0,\frac{\pi}{2})$ in a
fixed momentum direction. In the present truncation $\Delta$ would vanish for
$k_{ph}\rightarrow 0$, whereas $\kappa$ would remain nonzero.
In order to settle this issue more precisely one would have to follow
the flow of the Yukawa coupling that we neglected here.
For $t=0.3$ eV and $T=T_c/2$ we obtain $\Delta(0,\frac{\pi}{2})=7.7$ meV.

\begin{figure}[t]
\begin{center}
\begin{picture}(185,155)(20,-10)
\put(-5,76){$\eta$}
\put(110,-5){$T/T_c$}
\includegraphics[width=80mm,angle=0.]{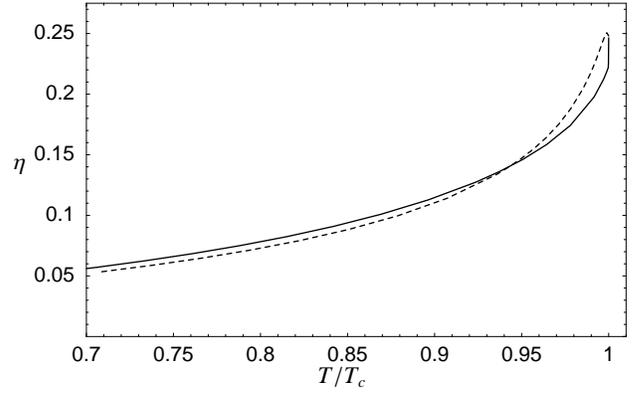}
\end{picture}
\vspace{-5mm}
\end{center}
\caption{Temperature dependence of the critical exponent $\eta$. The solid line is based on eq. (\ref{eq:impbeta}), the dashed line on eq. (\ref{eq:ansatz}).
}
\label{figetaT}
\end{figure}

We finally turn to the anomalous dimension for $T\leq T_c$. For weakly
inhomogeneous fields (low $\mathbf{q}^2$) the gradient terms in the
effective theory (Landau theory) for the composite bosons $\phi$ take
the form
\begin{eqnarray}\label{eq:La}
\Gamma=\frac{1}{T}\int_q \phi^*(q)A(\mathbf{q}^2,k_{ph})\,\mathbf{q}^2\phi(q)
\,.\end{eqnarray}
Our truncation (\ref{eq:U}) has not yet taken the
$\mathbf{q}^2$-dependence of $A$ into account: the flow equations
apply for $A(k)=A(\mathbf{q}^2=0,k)$. Nevertheless, the momentum
dependence of $A$ can be easily inferred by noting that $\mathbf{q}^2$
also acts as an infrared cutoff such that $R_k$ becomes ineffective
for $k^2<\mathbf{q}^2$. A reasonable approximation uses
$\partial_kA(\mathbf{q}^2,k)=\partial_kA(\mathbf{q}^2=0,k)\Theta(k^2-\mathbf{q}^2)$. The
solution at $k_{ph}$ reads for $\mathbf{q}^2>k^2_{ph}$,
$q=\sqrt{\mathbf{q}^2}$,
\begin{eqnarray}
A(q^2)=\bar{A}(q/\bar{q})^{-\eta}\,,\quad \eta=\int_{\ln q}^{\ln
\bar{q}}d\ln k\,\eta(k)/\ln(\bar{q}/q)
\,.\end{eqnarray}
In fig. \ref{figetaT} we show $\eta$ for $\bar{q}/q=e^5$ and observe
the characteristic dependence on $T$. As a result the static
correlation function $\langle \phi(\mathbf{r})\phi^*(0)\rangle_c$
decreases $\sim r^{-\eta}$ for large $r$. For a superconductor one
replaces in eq. (\ref{eq:La}) $\mathbf{q}^2$ by
$-(\nabla-2i|e|\mathbf{A})^2$ with $\mathbf{A}$ the space components
of the electromagnetic potential. Inserting $\phi=\phi_0$ this results
in an anomalous Landau theory for the magnetic field
$B=\partial_1A_2-\partial_2A_1$. Instead of the photon mass term $\sim
\mathbf{A}^2$ responsible for ordinary superconductivity in three
dimensions we now find a term $|\mathbf{A}|^{2-\eta}$. This may open a
window for a measurement of the anomalous dimension.

Acknowledgment:
H. C. Krahl thanks P. Strack and S. Diehl for useful discussions and comments.

\begin{thebibliography}{12}
\expandafter\ifx\csname natexlab\endcsname\relax\def\natexlab#1{#1}\fi
\expandafter\ifx\csname bibnamefont\endcsname\relax
  \def\bibnamefont#1{#1}\fi
\expandafter\ifx\csname bibfnamefont\endcsname\relax
  \def\bibfnamefont#1{#1}\fi
\expandafter\ifx\csname citenamefont\endcsname\relax
  \def\citenamefont#1{#1}\fi
\expandafter\ifx\csname url\endcsname\relax
  \def\url#1{\texttt{#1}}\fi
\expandafter\ifx\csname urlprefix\endcsname\relax\def\urlprefix{URL }\fi
\providecommand{\bibinfo}[2]{#2}
\providecommand{\eprint}[2][]{\url{#2}}

\bibitem{hubbard}
J. Hubbard,
  Proc. R. Soc. London, Ser. A 276 (1963).

\bibitem{anderson}
P. W. Anderson,
  Science \textbf{235}, 1196 (1987).

\bibitem{scalapino}
D. J. Scalapino,
  Physics Reports \textbf{250}, 329 (1995).

\bibitem{tsuei}
C. C. Tsuei, J. R. Kirtley,
  Rev.\ Mod. Phys. \textbf{72}, 969 (2000).

\bibitem{kosterlitzthouless73}
J. M. Kosterlitz and D. J. Thouless,
  J. Phys. C. \textbf{6}, 1181 (1973).

\bibitem{wetterichZ}
C. Wetterich,
  Z. Phys. C \textbf{57}, 451 (1993).

\bibitem{rufenacht}
A. Rufenacht, et al.,
  Phys. Rev. Lett. \textbf{96}, 227002 (2006).

\bibitem{corson}
J. Corson, et al.,
  Nature \textbf{398}, 221 (1999).

\bibitem{wang}
Y. Wang, et al.,
  Phys. Rev. B \textbf{73}, 024510 (2006).

\bibitem{lie}
L. Lie, et al.,
  Europhys. Lett. \textbf{72}, 451 (2005).

\bibitem{zanchi1}
D. Zanchi and H. J. Schulz,
  Z.Phys. \textbf{B103}, 339 (1997).

\bibitem{zanchi2}
D. Zanchi and H. J. Schulz,
  Europhys. Lett. \textbf{44}, 235 (1998).

\bibitem{halbothmetzner}
C. J. Halboth and W. Metzner,
  Phys.\ Rev. Lett. \textbf{85}, 5162 (2000).

\bibitem{halbothmetzner2}
C. J. Halboth and W. Metzner,
  Phys.\ Rev. B \textbf{61}, 7364 (2000).

\bibitem{salmhofer}
M. Salmhofer and C. Honerkamp,
  Prog. Theor. Phys. \textbf{105}, 1 (2001).

\bibitem{honerkamp01}
C. Honerkamp, M. Salmhofer, N. Furukawa and T. M. Rice,
  Phys. Rev. B \textbf{63}, 035109 (2001).

\bibitem{honerkampsalmhofer01}
C. Honerkamp and M. Salmhofer,
  Phys. Rev. Lett. \textbf{87}, 187004 (2001).

\bibitem{honerkamp02}
C. Honerkamp, M. Salmhofer and T. M. Rice,
  Eur. Phys. J. B \textbf{27}, 127 (2002).

\bibitem{katanin}
A. A. Katanin and A. P. Kampf,
  Phys. Rev. B \textbf{68}, 195101 (2003).



\bibitem{miyake}
K. Miyake, et al.,
  Phys.\ Rev. B \textbf{34}, 6554 (1986).

\bibitem{loh}
D. J. Scalapino, et al.,
  Phys.\ Rev. B \textbf{34}, 8190 (1986).

\bibitem{bickers}
N. E. Bickers, et al.,
  Int. J. Mod. Phys. B \textbf{1}, 687 (1987).

\bibitem{bbw04}
T. Baier, E. Bick and C. Wetterich,
  Phys.\ Rev. B \textbf{70}, 125111 (2004).

\bibitem{bbw05}
T. Baier, E. Bick and C. Wetterich,
  Phys.\ Rev. B \textbf{605}, 144 (2005).

\bibitem{berges_review02}
J. Berges, N. Tetradis and C. Wetterich,
  Phys. Rep. \textbf{363}, 223 (2002).

\bibitem{cw93}
C. Wetterich,
  Phys. Lett. B \textbf{301}, 90 (1993).

\bibitem{litim1}
D. F. Litim,
  Phys. Lett. B \textbf{486}, 92 (2000).

\bibitem{litim2}
D. F. Litim,
  Phys. Rev. D \textbf{64}, 105007 (2001).

\bibitem{vojta}
M. Vojta, Y. Zhang, S. Sachdev,
  Phys. Rev. Lett. \textbf{85}, 4940 (2000).

\bibitem{graeter95}
M. Graeter, C. Wetterich,
  Phys. Rev. Lett. \textbf{75}, 378 (1995).

\bibitem{vgersdorff}
G. von Gersdorff, C. Wetterich,
  Phys. Rev. B \textbf{64}, 054513 (2000).




















\end{thebibliography}


\end{document}